# Efficient Asymmetric Causality Tests

Abdulnasser Hatemi-J

Department of Economics and Finance, College of Business and Economics, UAE University

Email: AHatemi@uaeu.ac.ae

Abstract

Asymmetric causality tests are increasingly gaining popularity in different scientific fields. This approach corresponds better to reality since logical reasons behind asymmetric behavior exist and need to be considered in empirical investigations. Hatemi-J (2012) introduced the asymmetric causality tests via partial cumulative sums for positive and negative components of the variables operating within the vector autoregressive (VAR) model. However, since the residuals across the equations in the VAR model are not independent, the ordinary least squares method for estimating the parameters is not efficient. Additionally, asymmetric causality tests mean having different causal parameters (i.e., for positive or negative components), thus, it is crucial to assess not only if these causal parameters are individually statistically significant, but also if their difference is statistically significant. Consequently, tests of difference between estimated causal parameters should explicitly be conducted, which are neglected in the existing literature. The purpose of the current paper is to deal with these issues explicitly. An application is provided, and ten different hypotheses pertinent to the asymmetric causal interaction between two largest financial markets worldwide are efficiently tested within a multivariate setting.

# 1. Introduction

It is widely agreed that the cause and effect relationship has been one of the most important issues that has occupied minds since the dawn of humankind. However, what is meant by causality and how it should be measured is a polemical issue. In time series analysis the concept of causality originates from Wiener (1956). From his perspective causality is defined as "*For two simultaneously measured signals, if we can predict the first signal better by using the past information from the second one than by using the information without it, then we call the second signal causal to the first one.*" Granger (1969) introduced methods for testing the null hypothesis that the past values of a variable do not improve the forecast of another variable when all other relevant information is accounted for. This test is known as Granger causality test in literature. Since experimentation is normally not possible in economics, finance and other similar fields, Wiener's (1956) definition of causality implemented by Ganger's test is very useful for drawing casual inference based on empirical information extracted from non-experimental data. There are also alternative approaches for testing causality suggested by Sims (1972) and Geweke (1982), among others. Since the discovery of unit roots and its impact on the relationship between time series variables (Granger and Newbold, 1974; Dickey and Fuller, 1979), there have been additional developments in causality testing (Engle and Granger, 1987; Granger, 1986 and 1988, inter alia). Toda and Yamamoto (1995) introduced an approach which consists of adding one extra unrestricted lag of each variable in the VAR model in order to account for the impact of each unit root when tests for causality are conducted. This approach is based on asymptotic distributions. However, Hacker and Hatemi-J (2006) showed via simulations that Toda and Yamamoto (1995) approach has serious size distortions if the assumption of a normal distribution is not fulfilled and if the variance is not constant. Hacker and Hatemi-J (2006) offer a bootstrap version of the test with leverage adjustments that has better size properties. This bootstrap test was further improved by Hacker and Hatemi-J (2012) via endogenizing the selection of the optimal lag order in the bootstrap simulations, which improved both the size and the power properties of the bootstrap causality tests. Hatemi-J (2012) introduced asymmetric causality tests, which were made dynamic by Hatemi-J (2022) using an approach suggested by Phillips et. al., (2015). However, the previous asymmetric tests of Hatemi-J (2012) were conducted within the VAR model, which is not efficient in this case since the positive and negative components are not independent of each other. Furthermore, it is also important to test whether or not the deference between the causal parameters are significant jointly in addition to individual significant tests. Thus, the current paper aims at introducing efficient tests for asymmetric causality in a multivariate setting that can also be used for testing the significance of the difference in causal parameters pertinent to positive and negative innovations.[1] The methos is flexible and can include the multivariate GARCH effects with a multivariate t-distribution if needed.

The rest of this paper is structured as follows. Section 2 introduces the efficient tests for asymmetric causality and outlines a number of different hypotheses that can be interesting for testing within a joint multivariate setting. Section 3 offers an application, and the last section presents conclusions. An appendix is provided at the end of the paper.

2. **Methodology**

Consider the two time series variables $Z_1$ and $Z_2$ that are the focus of asymmetric causality. Each variable is assumed to integrated of the first degree with deterministic trend parts generated as the following:

$$Z_{1,t} = a + bt + Z_{1,t-1} + e_{i1,t} = at + \frac{t(t+1)}{2}b + Z_{1,0} + \sum_{j=1}^{t} e_{1,j} \qquad (1)$$

$$Z_{2,t} = c + dt + Z_{2,t-1} + e_{2,t} = ct + \frac{t(t+1)}{2}d + Z_{2,0} + \sum_{j=1}^{t} e_{2,j} \qquad (2)$$

For $t=1, \ldots, T$. Where $e_i$ is the white noise error term in each case for $i=1, 2$. The denotations $a$, $b$, $c$, $d$ represent the deterministic coefficients to be estimated. $Z_{i,0}$ represents the initial value for variable $i$. Granger and Yoon (2002) suggest to identify the positive and negative innovations as $e_{1,t}^+ := max(e_{1,t}, 0)$, $e_{2,t}^+ := max(e_{2,t}, 0)$, $e_{1,t}^- := min(e_{1,t}, 0)$ and $e_{2,t}^- := min(e_{2,t}, 0)$. Via these values, the partial cumulative sums of the variables, i.e. $Z_{1,t}^+$, $Z_{1,t}^-$, $Z_{2,t}^+$ and $Z_{2,t}^-$, can be constructed as the following:[2]

$$Z_{1,t}^+ = \rho \left( at + \left[\frac{t(t+1)}{2}\right]b + Z_{1,0} \right) + \sum_{j=1}^{t} e_{1,j}^+ \qquad (3)$$

$$Z_{1,t}^- = (1-\rho) \left( at + \left[\frac{t(t+1)}{2}\right]b + Z_{1,0} \right) + \sum_{j=1}^{t} e_{1,j}^- \qquad (4)$$

---

[1] For panel causality tests see Konya (2006), Emirmahmutoglu and Kose (2011) and Dumitrescu and Hurlin (2012), among others. Hatemi-J (2011, 2020) introduced asymmetric panel causality tests.

[2] These results were introduced by Hatemi-J (2014a) and proved by Hatemi-J and El-Khatib (2016). Statistical software components for transforming the data is provided by Hatemi-J (2014b) in Gauss, Hatemi-J and Mustafa (2016a) in programming language Visual Basic for Applications (VBA), and Hatemi-J and Mustafa (2016b) in Octave. Additional software component is provided by El-Khatib and Hatemi-J (2017) in C++.

$$Z_{2,t}^+ = q\left(at + \left[\frac{t(t+1)}{2}\right]b + Z_{1,0}\right) + \sum_{j=1}^{t} e_{2,j}^+ \tag{5}$$

$$Z_{2,t}^- = (1-q)\left(at + \left[\frac{t(t+1)}{2}\right]b + Z_{1,0}\right) + \sum_{j=1}^{t} e_{2,j}^- \tag{6}$$

Where $\rho$ is the proportion of the observations that represent positive changes for variable $Z_1$ and $q$ is the proportion of the observations that are positive changes for variable $Z_2$, which can used as wights for the deterministic trend part for each variables. This way of weighting is expected to be more precise compared to equal weighting for both components since the likelihood that the number of positive changes are equal to the number of negative changes are not usually confirm by real data for any variable. Note that the condition for correct transformation is fulfilled in each case because the sum of the positive and negative components results in obtaining each variable in its pristine format, that is, $Z_{1,t}^+ + Z_{1,t}^- = Z_{1,t}$ and $Z_{2,t}^+ + Z_{2,t}^- = Z_{2,t}$. These components can be used for conducting asymmetric causality tests. However, since the error terms across equations are not independent, the vector autoregressive (VAR) model that is usually estimated by the ordinary least squares (OLS) is not efficient despite being consistent. In order to implement asymmetric causality tests efficiently the following autoregressive seemingly unrelated regression equations (SURE), which has the same effect as the feasible generalized least squares (FGLS):[3]

$$\begin{bmatrix} Z_{1,t}^+ \\ Z_{2,t}^+ \\ Z_{1,t}^- \\ Z_{2,t}^- \end{bmatrix} = \begin{bmatrix} \lambda_1^+ \\ \lambda_2^+ \\ \lambda_1^- \\ \lambda_2^- \end{bmatrix} + \begin{bmatrix} \sum_{k^+=1}^{L^+} \beta_{1,k^+}^+ Z_{1,t-k^+}^+ + \sum_{k^+=1}^{L^+} \beta_{2,k^+}^+ Z_{2,t-k^+}^+ \\ \sum_{k^+=1}^{L^+} \gamma_{1,k^+}^+ Z_{1,t-k^+}^+ + \sum_{k^+=1}^{L^+} \gamma_{2,k^+}^+ Z_{2,t-k^+}^+ \\ \sum_{k^-=1}^{L^-} \beta_{1,k^-}^- Z_{1,t-k^-}^- + \sum_{k^-=1}^{L^-} \beta_{2,k^-}^- Z_{2,t-k^-}^- \\ \sum_{k^-=1}^{L^-} \gamma_{1,k^-}^- Z_{1,t-k^-}^- + \sum_{k^-=1}^{L^-} \gamma_{2,k^-}^- Z_{2,t-k^-}^- \end{bmatrix} + \begin{bmatrix} \varepsilon_{1,t}^+ \\ \varepsilon_{2,t}^+ \\ \varepsilon_{1,t}^- \\ \varepsilon_{2,t}^- \end{bmatrix} \tag{7}$$

---

[3] Sims (1980) introduced the VAR model, and the SURE model was pioneered by Zellner (1962).

Where $\varepsilon_{1,t}^+$, $\varepsilon_{2,t}^+$, $\varepsilon_{1,t}^-$ and $\varepsilon_{2,t}^-$ are the error terms, which can be dependent on each other. The optimal lag orders $L^+$ and $L^-$ can be determined via the minimization of an information criterion. The denotation $\lambda_1^+$ is the intercept for the equation of variable $Z_{1,t}^+$ and $\lambda_2^+$ is the intercept for the equation of $Z_{2,t}^+$. Likewise, $\lambda_1^-$ is the intercept for the equation of $Z_{1,t}^-$ and $\lambda_2^-$ represents the intercept for the equation that has $Z_{2,t}^-$ as the dependent variable. Denotations $\beta$ and $\gamma$ are parameters to be estimated for the lagged values of positive and negative components of the variables distinguished by the plus or minus signs. If the variance of the error terms are clustering, model (7) can also be estimated by the multivariate GARCH effect combined with a multivariate t-distribution if the data depicts fat tails.[4] A series of interesting null hypotheses are defined below, which can be tested within this multivariate setting. The null hypothesis that $Z_{2,t}^+$ does not cause $Z_{1,t}^+$ amounts to testing the following restrictions:

$$H_0: \sum_{k^+=1}^{P^+} \beta_{2,k^+}^+ = 0 \qquad (8)$$

The null hypothesis that $Z_{2,t}^-$ does not cause $Z_{1,t}^-$ means testing the following:

$$H_0: \sum_{k^-=1}^{P^-} \beta_{2,k^-}^- = 0 \qquad (9)$$

The null hypothesis that $Z_{2,t}^+$ or $Z_{2,t}^-$ do not cause $Z_{1,t}^+$ or $Z_{1,t}^-$ implies testing the following:

$$H_0: \sum_{k^+=1}^{P^+} \beta_{2,k^+}^+ = \sum_{k^-=1}^{P^-} \beta_{2,k^-}^- = 0 \qquad (10)$$

The null hypothesis that there is no asymmetric causality running from $Z_{2,t}$ on $Z_{1,t}$ means testing the following:

$$H_0: \sum_{k^+=1}^{P^+} \beta_{2,k^+}^+ = \sum_{k^-=1}^{P^-} \beta_{2,k^-}^- \qquad (11)$$

It should be mentioned that similar null hypotheses for the asymmetric causal impacts of $Z_{1,t}^+$ and/or $Z_{1,t}^-$ on $Z_{2,t}^+$ and/or $Z_{2,t}^-$ can be formulated. One can consider additional join hypotheses also. The null hypothesis that the two variables do not cause each other at all regardless of positive or negative changes is formulated as

$$H_0: \sum_{k^+=1}^{P^+} \beta_{2,k^+}^+ = \sum_{k^-=1}^{P^-} \beta_{2,k^-}^- = 0 \ and \ \sum_{k^+=1}^{P^+} \gamma_{1,k^+}^+ = \sum_{k^-=1}^{P^-} \gamma_{1,k^-}^- = 0 \quad (12)$$

---

[4] This approach is particularly useful if financial data is used. However, the issue can be determined empirically by conducting Hacker and Hatemi-J (2005) multivariate ARCH test. If the null hypothesis of no multivariate ARCH effects is rejected, then model (1) can be estimated by the generalized ARCH method. For a survey of this method see Silvennoinen and Teräsvirta (2009).

Finally, the null hypothesis that both variables do not cause each other asymmetrically is defined as the following:

$$H_0: \sum_{k^+=1}^{P^+} \beta_{2,k^+}^+ = \sum_{k^-=1}^{P^-} \beta_{2,k^-}^- \ and \ \sum_{k^+=1}^{P^+} \gamma_{1,k^+}^+ = \sum_{k^-=1}^{P^-} \gamma_{1,k^-}^- \quad (13)$$

Each null hypothesis can be tested via the Wald (1949) coefficient test, which is described in the appendix. It should be mentioned that the reverse hypotheses pertinent to the potential causal impact between the negative components can be formulated similarly. In addition, the dimension of the model can be increased by adding more variables. Since the variables have a unit root, an extra lag of each variable needs to be added into each equation in order to account for the impact of the unit root on causality tests according to Toda and Yamamoto (1995).[5]

## 3. An Application

The suggested asymmetric tests are applied to investigate the causal interaction between falling and rising prices in the two largest financial markets in the world. The monthly data for all share price indexes for the US and the Chinese markets are used for this purpose.[6] The period covers March of 1999 until May of 2024. The start of the sample is restricted by data availability for China. The source of the data is FRED database, which is provided online by the Federal Reserve Bank of St. Louis. The variables are expressed in natural logarithms before transforming them into positive and negative components using equations (3)-(6). It should be mentioned that only a drift was included in each case since there seems to be no need for the deterministic trend in any case based on Figures 1 and 2. The system of equations (1) is estimated subject to the multivariate GARCH(1, 1) effects combined with a multivariate t-distribution. The results of the conducted asymmetric causality tests are presented in Table 1. Ten different hypotheses are tested. The implication of each hypothesis is defined in Table 1. The order of the variables in the model is expressed as the following:

---

[5] A model for testing the asymmetric causal relationship between two variables is presented here. However, the dimension can be increased, and additional variables can be included.

[6] Testing for causality between international markets is referred to as testing for financial market integration in literature. This issue has important implication for investors, financial institutions, and policy makers. For a recent literature review on the topic see Patel et. al., (2022) and references therein.

$$\begin{bmatrix} Z_{1,t}^+ \\ Z_{2,t}^+ \\ Z_{1,t}^- \\ Z_{2,t}^- \end{bmatrix} = \begin{bmatrix} lnUS_t^+ \\ lnChina_t^+ \\ lnUS_t^- \\ lnChina_t^- \end{bmatrix}.$$

**Figure 1**: Time Plot of the All Shares Price Index for the US Market Expressed in Natural Logarithms.

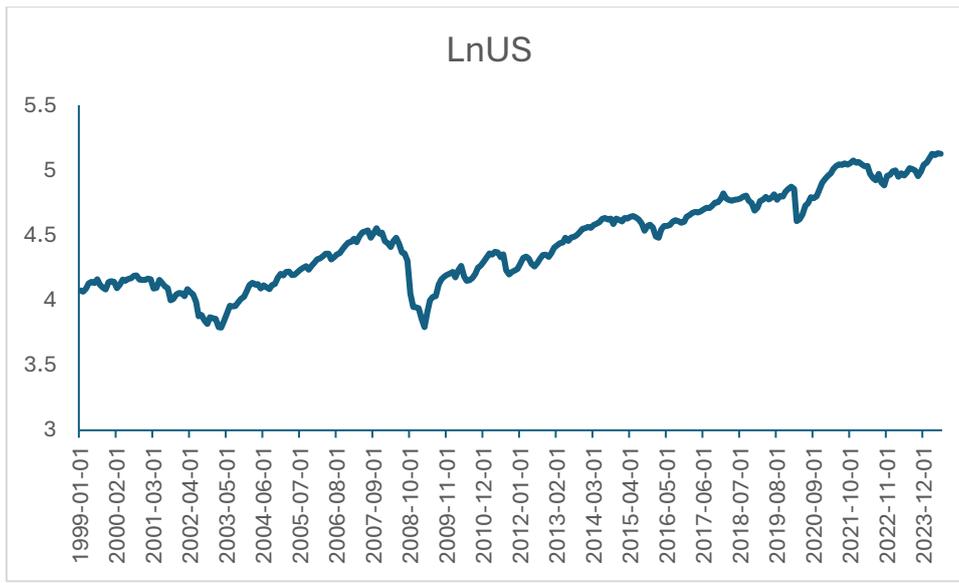

**Figure 2**: Time Plot of the All Shares Price Index for the Chinese Market Expressed in Natural Logarithms.

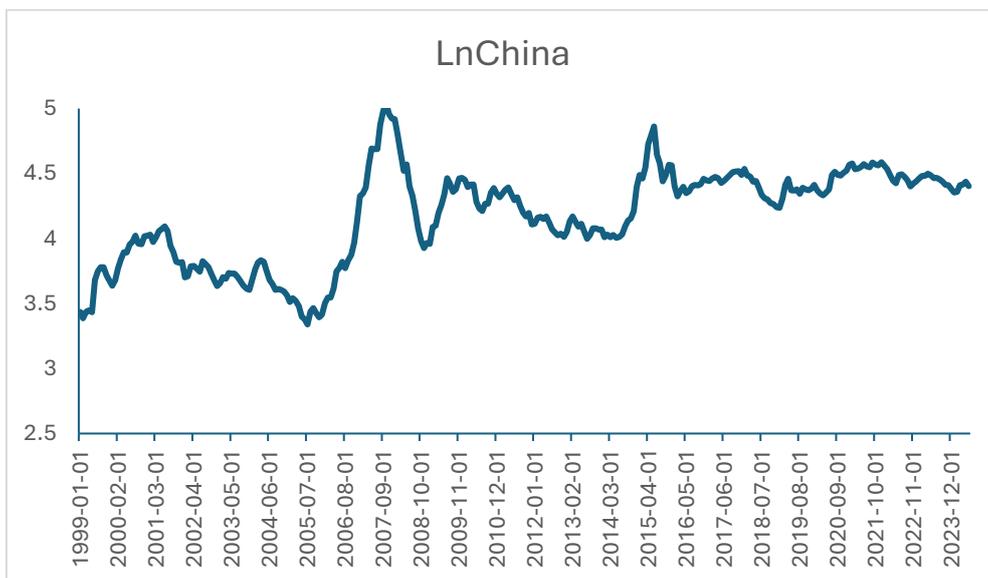

One lag of each variable was included in the model. However, an additional extra unrestricted lag was also added in the model to account for the unit root impact as per recommendations of Toda and Yamamoto (1995). The results show that the US, and the Chinese financial markets are interacting asymmetrically in causal terms. A price decrease in the US market results in a price decrease in the Chenese market and *vice versa*. The same can be said about a price decrease. Of the ten relevant null hypotheses of no causality that are tested only one could not be rejected (i.e., the individual null hypothesis that a falling market in the US does not cause a falling market in China, even though the joint hypothesis is rejected).

The estimated parameters reveal further that the falling market in China affects the falling US market more than the reverse impact (i.e., 0.713761 compared to 0.019351). While a rising market in the US has a stronger causal impact on the rising Chinese market compared to the reverse causal impact (i.e., 0.151631 compared to 0.025263).

**Table 1**. Estimation Results of the Efficient Asymmetric Causality Tests.

| Parameters | Estimated Parameters | |
|---|---|---|
| $\beta_{2,1}^+$ | 0.025263 | |
| $\beta_{2,1}^-$ | 0.713761 | |
| $\gamma_{1,1}^+$ | 0.151631 | |
| $\gamma_{1,1}^-$ | 0.019351 | |
| **Null Hypothesis** | **P-value** | **Implication of the Null Hypothesis** |
| $H_0: \beta_{2,1}^+ = 0$ | < 0.00001 | A rising market in China does not cause a rising market in US. |
| $H_0: \beta_{2,1}^-$ | < 0.00001 | A falling market in China does not cause a falling market in US. |
| $H_0: \beta_{2,1}^+ = 0 \ and \ \beta_{2,1}^- = 0$ | < 0.00001 | Neither rising nor falling markets in China cause rising or falling markets in US. |
| $H_0: \beta_{2,1}^+ - \beta_{2,1}^- = 0$ | 0.00010 | The impact of rising and falling markets in China is the same on US markets (symmetric causality). |
| $H_0: \gamma_{1,1}^+ = 0$ | 0.00780 | A rising market in US does not cause a rising market in China. |
| $H_0: \gamma_{1,1}^-$ | 0.12500 | A falling market in US does not cause a falling market in China. |
| $H_0: \gamma_{1,1}^+ = 0 \ and \ \gamma_{1,1}^- = 0$ | 0.00720 | Neither rising nor falling markets in US cause rising or falling markets in China. |
| $H_0: \gamma_{1,1}^+ - \gamma_{1,1}^- = 0$ | 0.02500 | The impact of rising and falling markets in US is the same on China markets (symmetric causality). |
| $H_0: \beta_{2,1}^+ = 0, \ \beta_{2,1}^- = 0, \ \gamma_{1,1}^+ = 0 \ and \ \gamma_{1,1}^- = 0$ | < 0.00001 | These two markets are not causing each other for both rising and falling markets. |
| $H_0: \beta_{2,1}^+ - \beta_{2,1}^- = 0$ and $\gamma_{1,1}^+ - \gamma_{1,1}^- = 0$ | < 0.00001 | The joint causal impacts of the two markets on each other are symmetric. |

Notes: The multivariate GARCH parameters are not presented here. However, most of these 30 estimated parameters are statistically significant. An unrestricted lag of each variable was included in the model for accounting for the impact of the unit root based on Toda and Yamamoto (1995) results.

## 4. Conclusions

Testing for asymmetric causality is increasingly gaining popularity when time series or panel data are used in different scientific disciplines. This approach accords better with reality since there are logical reasons behind asymmetric behavior that need to be considered in empirical investigations. Hatemi-J (2012) introduced the asymmetric causality tests that are based on partial cumulative sums for positive and negative components of the variables within the vector autoregressive (VAR) model. The critique of this approach is that the positive and negative components are not independent and therefore the ordinary least squares method used for estimating the parameters in the VAR model is not efficient in this case. In addition, since the proposed causality tests recommend that there can be different causal parameters (i.e., for positive or negative components of each variable), it is necessary to assess not only if these causal parameters are individually statistically significant, but also if their difference is statistically significant. Hence, tests of difference between estimated causal parameters should explicitly be conducted. The current paper deals with these issues and present a system of equations for conducting the tests. The system can be estimated efficiently, and it can also account for the multivariate GARCH effects combined with the multivariate t-distribution accounting for potential fat-tails effects. An application is provided, and ten different hypotheses are tested that are pertinent to the asymmetric causal interaction between the two largest financial markets in the world. The results show that these two financial markets are causally related and the causal impact of each market on the other is asymmetric. These asymmetric causal impacts between the markets are statistically significant. Furthermore, the estimations uncover that the impact of falling markets in China on the falling US markets is greater compared to the reverse impact. However, the rising markets in the US have a stronger causal impact on the rising Chinese markets contrasted with the reverse causal effect. These results can be useful to investors, financial institutions, and policy makers.

Futures studies can discover the extent of the potential asymmetric causal interaction between different variables or assets for different regions and across different time horizons by applying the suggested efficient asymmetric tests.

**Acknowledgements**

The article is supported financially by the CBE Annual Research Program (CARP) 2024 grant of the United Arab Emirates University. The common disclaimer applies, nevertheless.

**References**

Dickey D.A. and Fuller W.A. (1979) Distribution of the estimators for autoregressive time series with a unit root, *Journal of the American Statistical Association*, 74(366a), 427-431.

Dumitrescu E.I. and Hurlin C. (2012) Testing for Granger non-causality in heterogeneous panels, *Economic modelling*, 29(4), 1450-1460.

El-Khatib Y. and Hatemi-J A. (2017) ASYM_CAUS: C++ Module for Transforming an Integrated Variable with Deterministic Trend Parts into Negative and Positive Cumulative Partial Sums, *Statistical Software Components CPP001*, Boston College Department of Economics. Available from https://ideas.repec.org/c/boc/bocode/cpp001.html

Emirmahmutoglu F. and Kose N. (2011) Testing for Granger causality in heterogeneous mixed panels, *Economic Modelling*, 28(3), 870-876.

Engle R.F. and Granger C.W. (1987) Co-integration and error correction: representation, estimation, and testing, *Econometrica*, 251-276.

Geweke J. (1982) Measurement of linear dependence and feedback between multiple time series, *Journal of the American Statistical Association*, 77(378), 304-313.

Granger, C.W.J. (1969) Investigating Causal Relations by Econometric Models and Cross-spectral Methods, *Econometrica,* 37, 424-439.

Granger C.W.J. (1986) Developments in the Study of Cointegrated Economic Variables, *Bulletin of Economics and Statistics*, 48(3), 213-28.

Granger C.W.J. (1988) Causality, cointegration, and control, *Journal of Economic Dynamics and Control*, 12, 551-559.

Granger C.W.J. and Yoon G. (2002) Hidden Cointegration. Department of Economics Working Paper. University of California. San Diego.

Granger C.W.J. and Newbold P. (1974) Spurious Regressions in Econometrics. *Journal of Econometrics*, 2, 111-20.

Hacker R. and Hatemi-J A. (2005) A test for multivariate ARCH effects, *Applied Economics Letters*, 12(7), 411-417.

Hacker R. S. and Hatemi-J A. (2006) Tests for causality between integrated variables using asymptotic and bootstrap distributions: theory and application, *Applied Economics*, 38(13), 1489-1500.

Hacker S. and Hatemi-J A. (2012) A bootstrap test for causality with endogenous lag length choice: theory and application in finance, *Journal of Economic Studies*, 39(2), 144-160.

Hatemi-J, A (2012) Asymmetric Causality Tests with an Application, *Empirical Economics*, 43(1), 447-456.

Hatemi-J A. (2014a) Asymmetric Generalized Impulse Response Functions with an Application in Finance, *Economic Modelling*, 36(C), 18-22.

Hatemi-J A. (2014b) ASCOMP: GAUSS module to Transform Data into Cumulative Positive and Negative Components, *Statistical Software Components G00015*, Boston College Department of Economics. Available from
https://ideas.repec.org/c/boc/bocode/g00015.html

Hatemi-J A. (2020) Asymmetric Panel Causality Tests with an Application to the Impact of Fiscal Policy on Economic Performance in Scandinavia, *Economia Internazionale/International Economics*, 73(3), 389-404.

Hatemi-J A. (2022) Dynamic asymmetric causality tests with an application, *Engineering Proceedings*, 18(1), 41.

Hatemi-J A. and El-Khatib Y. (2016) An extension of the asymmetric causality tests for dealing with deterministic trend components, *Applied Economics*, 48(42), 4033-4041.

Hatemi-J A. and Mustafa A. (2016a) A MS-Excel Module to Transform an Integrated Variable into Cumulative Partial Sums for Negative and Positive Components with and without Deterministic Trend Parts, MPRA Paper 73813, University Library of Munich, Germany.

Hatemi-J A. and Mustafa A. (2016b) TDICPS: OCTAVE module to Transform an Integrated Variable into Cumulative Partial Sums for Negative and Positive Components with Deterministic Trend Parts, *Statistical Software Components OCT001*, Boston College Department of Economics. Available from
https://ideas.repec.org/c/boc/bocode/oct001.html

Kónya L. (2006) Exports and growth: Granger causality analysis on OECD countries with a panel data approach, *Economic Modelling*, 23(6), 978-992.

Patel R., Goodell J.W., Oriani M.E., Paltrinieri, A. and Yarovaya L. (2022) A bibliometric review of financial market integration literature, *International Review of Financial Analysis*, 80, 102035.

Phillips P.C., Shi S. and Yu J. (2015) Testing for multiple bubbles: Historical episodes of exuberance and collapse in the S&P 500, *International Economic Review*, 56(4), 1043-1078.

Silvennoinen A. and Teräsvirta T. (2009) Multivariate GARCH models. In Handbook of financial time series (pp. 201-229). Berlin, Heidelberg: Springer Berlin Heidelberg.

Sims C.A. (1972) Money, income, and causality, *American Economic Review*, 62(4), 540-552.

Sims C.A. (1980) Macroeconomics and Reality, *Econometrica*, 48(1), 1-48.

Toda H.Y. and Yamamoto T. (1995) Statistical Inference in Vector Autoregressions with Possibly Integrated Processes, *Journal of Econometrics,* 66, 225-250.

Wald A. (1949) Statistical decision functions. *Annals of Mathematical Statistics*, 20(2), 165-205.

Wiener N. (1956) The theory of prediction. In: Beckenbach, E. (Ed.), Modern Mathematics for Engineers. McGraw-Hill, New York.

Zellner A (1962) An Efficient Method of Estimating Seemingly Unrelated Regressions and Tests for Aggregation Bias, *Journal of the American Statistical Association*, 57, 348-368.

## Appendix

Via certain denotations, model (1) can be expressed in a general format as the following:

$$X = \begin{bmatrix} X_1 \\ \vdots \\ X_n \end{bmatrix} = \begin{bmatrix} Z_1 & \cdots & 0 \\ \vdots & \ddots & \vdots \\ 0 & \cdots & Z_n \end{bmatrix} \times \begin{bmatrix} C_1 \\ \vdots \\ C_n \end{bmatrix} + \begin{bmatrix} \varepsilon_1 \\ \vdots \\ \varepsilon_n \end{bmatrix} = ZC + \varepsilon \qquad (A1)$$

Here $X_i$ is a $n \times 1$ vector of the dependent variables and $Z_i$ is $n \times (n \times P)$ the matrix of independent variables. Note that $n$ is the number of variables for each unit, $P$ is the lag order. The variance-covariance matrix of the error terms $\varepsilon_i$ for $i = 1, \ldots, n$ is denoted by $\Omega$ and it is expressed as the following:

$$\Omega = \begin{bmatrix} \sigma_{11} & \cdots & \sigma_{1n} \\ \vdots & \ddots & \vdots \\ \sigma_{n1} & \cdots & \sigma_{nn} \end{bmatrix} \qquad (A2)$$

The efficient value of *C* can be estimated via the generalized least squares method as the following:

$$\hat{C} = [Z'(\Omega^{-1} \otimes I)Z]^{-1} Z'(\Omega^{-1} \otimes I)X \qquad (A3)$$

The denotation $\otimes$ is the Kronecker operator while *I* represents the identity matrix. The null hypothesis of no causality in this context amounts to the following expression:

$$H_0: R\hat{C} = 0,$$

$R$ is a matrix consisting of one and zero elements for indicating which parameters should be restricted to zero when the null hypothesis is tested. This null hypothesis can be tested by the multivariate Wald (1949) coefficient test, which is formulated as

$$Wald = (R\hat{C})'[R\widehat{Var}(\hat{C})R']^{-1}(R\hat{C}) \qquad (A4)$$

The symbol $\widehat{Var}(\hat{C})$ signifies the estimated variance-covariance matrix of the estimated regression parameters. Conditional on the assumption of normality, the Wald test has a chi-square distribution with the number of restrictions imposed by the null hypothesis as the degrees of freedom.